\documentclass[%
 aip,
 jmp,%
 amsmath,amssymb,
 reprint,%
]{revtex4-1}

\usepackage{subfigure,amsmath,graphicx,epsfig,amssymb,float}
\usepackage{dcolumn}
\usepackage{bm}
\usepackage{epstopdf}

\begin{document}

\preprint{AIP/123-QED}

\title{Hubbard Physics in the PAW GW Approximation}

\author{J. M. Booth}
\email{jamie.booth@rmit.edu.au}
\affiliation{ 
Theoretical Chemical and Quantum Physics, School of Science, RMIT University, Melbourne, VIC, Australia}%

\author{D. W. Drumm}%
\affiliation{ 
Theoretical Chemical and Quantum Physics, School of Science, RMIT University, Melbourne, VIC, Australia}%
\affiliation{Australian Research Council Centre of Excellence for Nanoscale BioPhotonics}

\author{P. S. Casey}
\affiliation{
CSIRO Manufacturing and Materials, Clayton, VIC, Australia
}%
\author{J. S. Smith}
\affiliation{ 
Theoretical Chemical and Quantum Physics, School of Science, RMIT University, Melbourne, VIC, Australia}%
\author{S. P. Russo}
\affiliation{ 
Theoretical Chemical and Quantum Physics, School of Science, RMIT University, Melbourne, VIC, Australia}%

\date{\today}

\begin{abstract}
It is demonstrated that the signatures of the Hubbard Model in the strongly interacting regime can be simulated by modifying the screening in the limit of zero wavevector in Projector-Augmented Wave GW calculations for systems without significant nesting. This modification, when applied to the Mott insulator CuO, results in the opening of the Mott gap by the splitting of states at the Fermi level into upper and lower Hubbard bands, and exhibits a giant transfer of spectral weight upon electron doping. The method is also employed to clearly illustrate that the M$_{1}$ and M$_{2}$ forms of vanadium dioxide are fundamentally different types of insulator. Standard GW calculations are sufficient to open a gap in M$_{1}$ VO$_{2}$, which arise from the Peierls pairings filling the valence band, creating homopolar bonds. The valence band wavefunctions are stabilized with respect to the conduction band, reducing polarizability and pushing the conduction band eigenvalues to higher energy. The M$_{2}$ structure however opens a gap from strong on-site interactions; it is a Mott insulator.
\end{abstract}

\maketitle

\section{\label{sec:level1}Introduction}

Strongly correlated electrons in transition metal oxides generate phenomena such as high temperature superconductivity, colossal magnetoresistance and metal-insulator transitions, which offer enormous potential for new generations of devices. \cite{Dagotto2005,Takagi2010}  The development of density functional theory (DFT)\cite{Hohenberg1964} and its applications to weakly correlated materials such as the $p$-block semiconductors has significantly facilitated material design. Strongly correlated materials however, have not received such benefits as accurate approximations to the quantum many-body problem have proven elusive. \cite{Imada1998}  

The paradigm for the description of strong electron correlations is the Hubbard Hamiltonian \cite{Hubbard1963} (equation 1), which exhibits a competition between hopping, given by the $t$ term, and repulsion, given by the $U$ term, which lies at the heart of strongly correlated systems. If the orbitals between which the electrons hop are localized, such as transition metal d- or f-orbitals, then hopping to an already occupied site incurs an energy penalty, $U$. 
\begin{equation}
H=-t\sum\limits_{\langle ij\rangle}^{}(c^{\dagger}_{i\sigma}c_{j\sigma}+c^{\dagger}_{j\sigma}c_{i\sigma})+\\U\sum\limits_{i}^{}n_{i\uparrow}n_{i\downarrow}
\end{equation}

Therefore, at half-filling a tendency towards sites being singly occupied, and thus insulating behavior is observed. This localization is a many-body effect: the $U$ term is incurred by electrons encountering each other, and therefore the localization of each electron requires knowledge of where the other electrons are. Since the discovery of high temperature superconductivity, attempts to develop \textit{ab initio} tools to describe this have taken on increased urgency. To-date only two approaches have emerged which have been able to generate significant insight. 

Dynamical Mean Field Theory,\cite{Kotliar2006} in its DFT+DMFT form takes input wavefunctions from Density Functional Theory for the bands of interest, then applies sophisticated Monte Carlo approaches to the interactions of these electrons via the model Hubbard Hamiltonian. DMFT has provided significant insight into the nature of some strongly correlated systems such as V$_{2}$O$_{3}$,\cite{Mo2003} LaO$_{1-x}$F$_{x}$FeAs, and FeSe.\cite{Aichhorn2010} The technique revolves around isolation of the bands of interest, and projecting them into real space, usually onto Wannier functions,\cite{Lechermann2006} which are a natural basis for the Hubbard Model (indeed, this was the basis chosen by Hubbard himself). Contributions to the electron self-energy from screening and scattering are evaluated using a Mean Field Approximation based on the Anderson Impurity Model, or cluster variations.\cite{Kotliar2006} Despite such successes, DMFT and its variations require considerable power and sophistication when applied to real systems. Other complementary techniques therefore become attractive when the number of correlated bands in the system is large.

The GW approximation \cite{Hedin1965,Aryasetiawan1998} is a many-body perturbation theory approach which in the last decade has been integrated with density functional methods to provide accurate \textit{ab initio} calculations of the electronic structures of materials such as Si, GaAs etc.\cite{Faleev2004,Shishkin2007} This approach takes the bare Hartree-Fock interaction, well known to overstate the interactions between electrons, and screens it with the dielectric matrix, usually calculated in the Random Phase Approximation (RPA). This approach has exhibited significant improvements over DFT, and is now a standard component of most \textit{ab initio} packages. However, to maintain computational tractability, the self-energy is constructed using non-interacting Green functions and completely neglects electron scattering diagrams, and thus significant modification is required to apply the GW Approximation to Mott insulating systems. 

By far the most common method of ``correcting" the RPA-based GW method is to modify the input wavefunctions and eigenvalues by calculating them using either the DFT+U approach of Anisimov \textit{et al.},\cite{Anisimov1997} or with hybrid functionals.\cite{Heyd2003a} Jiang \textit{et al.} employed the LDA+U method to the electronic structures of oxides of the lanthanides\cite{Jiang2009} and the first row transition metal oxides MnO, FeO, CoO and NiO.\cite{Jiang2010} Considerable improvement was found in the band gaps in both studies.   

R{\"o}dl \textit{et al.}\cite{Claudia2015} compared self-consistent GW calculations using input wavefunctions from DFT+U and hybrid functional calculations utilizing different iteration schemes for the calculation of the photoemission spectrum of CuO. Of these, the authors found that self-consistency in both eigenvalues and wavefunctions vastly overestimated the band gap if the static screening was first approximated by the HSE06 functional,\cite{Heyd2003a} or PBE\cite{Perdew1996}+U calculations. The authors found that an approach in which a screened Coulomb interaction in good agreement with experiment was used and held fixed achieved the best approximation to experimental data. 

Lany\cite{Lany2013a} introduced an arbitrary (attractive) on-site addition to the local potential to GW calculations of 3$d$ metal oxides (including CuO) in an attempt to both obtain band gaps with better agreement with experiment, and fix the incorrect band ordering generated by using hybrid functional input to GW calculations. Again considerable improvement was found for band gaps and band ordering. We discuss the basis of the DFT+U and hybrid functional input approach in comparison to the method developed in this work in Section IIC. 

Gatti and Guzzo\cite{Gatti2013} applied the ``GW+C" method (the C stands for Cumulant), which is obtained from a decoupling of the elements of the exact one-electron Green function in the Dyson equation,\cite{Guzzo2011} to the study the effect of satellites on the electronic structure of SrVO$_{3}$. This approach successfully renormalized the V $3d$ bands and satellites near the Fermi level from first principles. This renormalization gave good agreement with experiment, without the use of model Hamiltonians.

Recently, some attempts to apply GW to strongly correlated systems have focused on combining it with DMFT, however the significant theoretical and computational complexity has so far limited this approach to model Hamiltonians,\cite{Ayral2012,Ayral2013} adatoms on surfaces,\cite{Hansmann2013} and SrVO$_{3}$.\cite{Tomczak2012,Sakuma2013}

In this work, we take a different approach. Rather than adjust the input wavefunctions, we approximate the effect of on-site repulsion by partially unscreening the Coulomb interaction in the limit of low wavevector in the GW calculation. This approach simulates the scattering resulting from the Hubbard $U$ term by mimicking the closing of the polarization bubbles by electrons on other sites. We refer to this technique as ``Partially Screened GW," or PS-GW.

 \begin{figure*}
   \includegraphics[width=1.6\columnwidth]{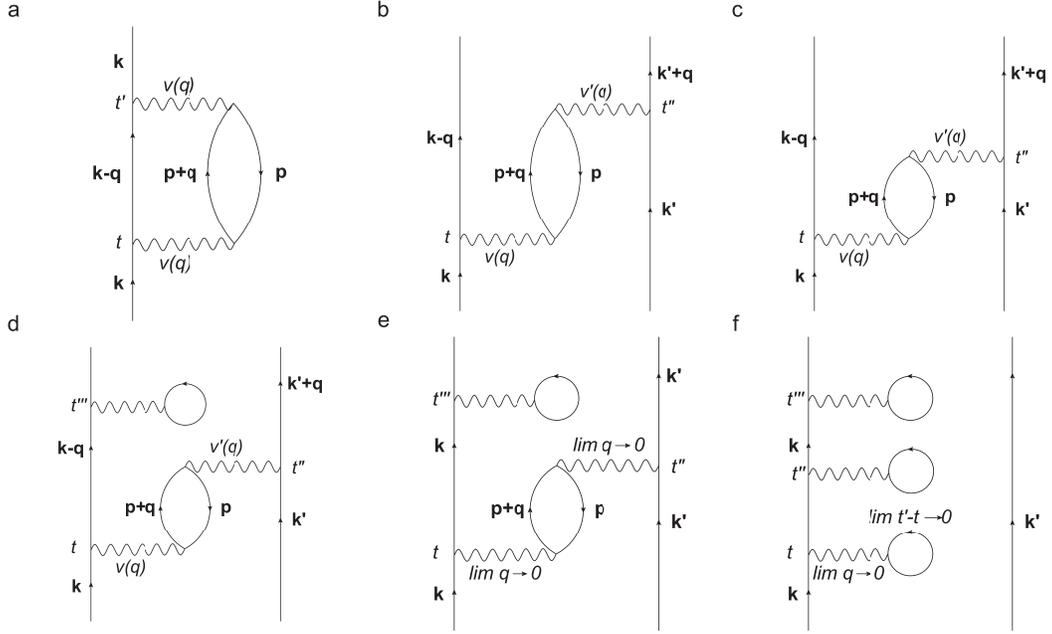}
   \caption{Feynman diagrams of relevant interactions in the systems studied, a) Polarization bubble of the type summed over to generate the screened interaction in the Random Phase Approximation, b) scattering vertex $\Gamma(\mathbf{k},\mathbf{k^{\prime}})$ resulting from the interaction of the polarization bubble with another momentum state, c) as per b) but with a large overlap shifting the characteristic frequency of the bubble to higher $\omega$, d) scattering vertex with Hartree interaction at later time, e) low $\mathbf{q}$ limit of the vertex and f) low $\mathbf{q}$ limit, and limit as $\omega \rightarrow \infty$ of the vertex.}
 \end{figure*}

 \begin{figure*}
   \includegraphics[width=2.0\columnwidth]{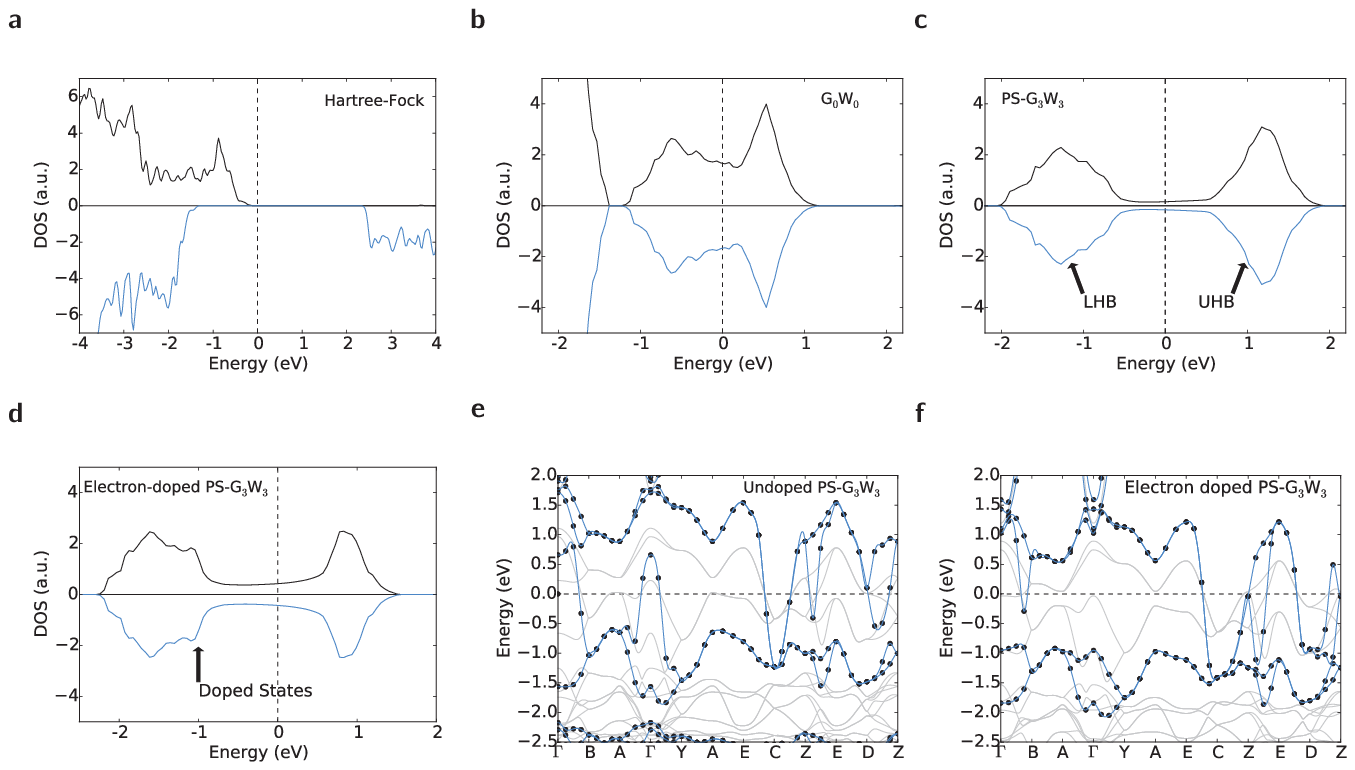}
   \caption{a) Hartree-Fock calculation of CuO b) Unmodified frequency-dependent spin-resolved $G_{0}W_{0}$ calculation of the states near E$_{F}$ of CuO, c) static PS-G$_{3}$W$_{3}$ calculation of CuO, d) static PS-G$_{3}$W$_{3}$ calculation of CuO doped with one electron, e) DFT (gray lines) and static PS-G$_{3}$W$_{3}$ (black filled circles) band structures of CuO, f) DFT (gray lines) and static PS-G$_{3}$W$_{3}$ (black filled circles) band structures of electron doped CuO. PS-GW derived eigenvalues are fitted with blue splines as a guide to the eye.}
 \end{figure*}

\section{\label{sec:level1}Methods}
\subsection{Assumptions}
The approach taken in this work is based around the following assumptions: i) the GW Approximation based on the screened interaction calculated in the RPA gets the electronic structure ``mostly" correct. This assumption infers that RPA-based GW calculations can be ``corrected" for the effects of scattering in the self-energy, ii) the most significant contributions to the self-energy of electrons near the Fermi level ($E_{F}$), come from inter-site hopping in the systems studied in this work, iii) these transitions manifest in the low $\mathbf{q}$ limit of the dielectric response, and thus the transitions which create double occupancies manifest in this limit, iv) correlations will heavily suppress transitions for high frequencies, such that the static limit of $\chi^{0}(\mathbf{q}, \omega)$ is a more accurate representation of the response of real systems. 

Of these three assumptions, ii) and iv) are the most difficult to justify. Assumption iii) follows from ii) and is supported by calculations; if ii) holds then calculating $\epsilon(\mathbf{q},\omega)$ on different $\mathbf{q}$-point grids should give the same results, and this is indeed the case. For all systems studied in this work, the dielectric response only depends  on the low $\mathbf{q}$ transitions. Changing the number of $\mathbf{q}$ points leaves the response invariant. Justification of assumption i) follows from the data itself. If RPA-GW can be corrected (using sound theoretical arguments) such that agreement with experimental results is achieved, then the approach is justified. We take the same approach to the justification of assumption iv), which has the added benefit of significantly reducing the computational resources required.

Assumption ii) depends to a large extent on the method used to generate the input wavefunctions. The two Mott systems studied here, CuO and M$_{2}$ VO$_{2}$, are both Monoclinic. This Monoclinic structure results from the adoption of a charge density wave which stabilizes the oxygen states, reducing their energies with respect to the metal $d$-states at the Fermi level. This stabilization arises from a change in nuclear potential, and is thus well reproduced by DFT. For these systems, DFT provides an adequate starting point for the method described here. For other systems, such as Cu$_{2}$O,\cite{Lany2013a} this is not the case, and the method used to generate the input wavefunctions must be modified accordingly, such as by employing the DFT+U method,\cite{Anisimov1997} or Hybrid Functionals.\cite{Heyd2003}

Thus the overall premise of this work is that: in \textit{ab initio} calculations of Mott systems, the creation of double occupancies results from the low $\mathbf{q}$ limit of the dielectric matrix, and that modification of the RPA screening in this limit can reproduce the signatures of Mott systems.

\subsection{GW Calculations}
The GW approximation is encapsulated by the Hedin equations,\cite{Hedin1965} which constitute a self-consistent approach to evaluating the electron self-energy: 
\begin{eqnarray}
\chi^{0}=-iGG\Gamma\\
\Gamma=1+\frac{\delta\Sigma}{\delta G}GG\Gamma\\
W=\epsilon^{-1}\nu\\
\epsilon=1-\nu\chi^{0}\\
\Sigma=iGW\Gamma
\end{eqnarray}
where $G$ is a single particle Green function, $\chi^{0}$ is the irreducible polarizability, $\Gamma$ is a Vertex Function, $W$ is the screened interaction, $\epsilon$ is the quantum dielectric matrix, and $\Sigma$ is the self-energy. In \textit{ab initio} calculations, $\Sigma$ is used in place of the DFT exchange-correlation energy in a ``quasiparticle" Hamiltonian to generate band eigenvalues.\cite{Shishkin2006} The polarizability matrix, $\chi=-iGG\Gamma$, is usually evaluated by setting $\Gamma$=1 (unless excitonic effects are important, in which case $\Gamma$ is approximated by solving the Bethe-Salpeter equation\cite{Schmidt2003}), and the Green functions used are non-interacting.\cite{Hybertsen1987,Gajdos2006,Shishkin2006} Standard perturbation theory provides a computational form of this independent-particle polarizability matrix weighted by the Hartree-Fock overlap integrals for the vertex in reciprocal space, ($\chi^{0}(\omega)$) as per:\cite{Hybertsen1987}
\begin{multline}
\chi^{0}(\mathbf{q},\omega)=\frac{1}{\Omega}\sum\limits_{\mathbf{G}\mathbf{G}^{\prime}}^{}\sum\limits_{nn^{\prime}}^{}2w_{\mathbf{k}}(f_{n^{\prime}\mathbf{p+q}}-f_{n\mathbf{p}})\\\times
\frac{\langle\psi_{n^{'}\mathbf{p+q}}|e^{-i\mathbf{(q+G)r}}|\psi_{n\mathbf{p}}\rangle\langle\psi_{n\mathbf{p}}|e^{i\mathbf{(q+G^\prime)r^\prime}}|\psi_{n^{'}\mathbf{p+q}}\rangle}{\omega+\epsilon_{n^{\prime}{\mathbf{p}}+\mathbf{q}}-\epsilon_{n\mathbf{p}}+i\eta \text{sgn}[\epsilon_{n\mathbf{p}}-\epsilon_{n^{\prime}\mathbf{p}+\mathbf{q}}]}
\end{multline}
In the VASP implementation of the GW approximation used in this study, the polarizability is combined with the Coulomb interaction into the dielectric matrix (equation (5)) as per:\cite{Shishkin2006}
\begin{equation}
\epsilon_{\mathbf{q}}(\mathbf{G},\mathbf{G^{\prime}},\omega)=\delta_{\mathbf{G},\mathbf{G^{\prime}}}-\frac{4\pi e^{2}}{|\mathbf{q}+\mathbf{G}||\mathbf{q}+\mathbf{G^{\prime}}|}\chi^{0}_{\mathbf{q}}(\mathbf{G},\mathbf{G^{\prime}},\omega)
\end{equation}

This dielectric matrix screens the bare Coulomb interaction to generate the screened interaction (equation (4)):\cite{Shishkin2006}
\begin{equation}
W_{\mathbf{q}}(\mathbf{G},\mathbf{G^{\prime}},\omega)=4\pi e^{2}\frac{1}{|\mathbf{q}+\mathbf{G}|}\epsilon^{-1}_{\mathbf{q}}(\mathbf{G},\mathbf{G^{\prime}},\omega)\frac{1}{|\mathbf{q}+\mathbf{G^{\prime}}|}
\end{equation}
The computational version of equation (6) is then given by:\cite{Shishkin2006}
\begin{multline}
\bar{\Sigma}(\omega)_{n\mathbf{k}n\mathbf{k}}=\frac{1}{\Omega}\sum_{\mathbf{qGG^{\prime}}}^{}\sum_{n^{\prime}}^{}\frac{i}{2\pi}\int_{0}^{\infty}d\omega^{\prime}\bar{W}_{\mathbf{q}}(\mathbf{G},\mathbf{G^{\prime}},\omega^{\prime})\\\times\langle\psi_{n\mathbf{k}}|e^{-i\mathbf{(q+G)r}}|\psi_{n^{\prime}\mathbf{k-q}}\rangle\langle\psi_{n^{\prime}\mathbf{k-q}}|e^{i\mathbf{(q+G^\prime)r^\prime}}|\psi_{n\mathbf{k}}\rangle\\\times\bigl(\frac{1}{\omega+\omega^{\prime}-\epsilon_{n^{\prime}\mathbf{k-q}}+i\eta\text{sgn}[\epsilon_{n^{\prime}\mathbf{k-q}}-\mu]}\\-\frac{1}{\omega-\omega^{\prime}-\epsilon_{n^{\prime}\mathbf{k-q}}+i\eta \text{sgn}[\epsilon_{n^{\prime}\mathbf{k-q}}-\mu]}\bigr)\
\end{multline}

Thus the two vertices of the screened interaction are split, with the photon propagator terms appearing in equations (8) and (9), while the Hartree-Fock overlap integrals are the angle-bracket terms contained in equations (7) and (10). In the VASP implementation of the GW approximation the bare Coulomb kernel is subtracted from the screened interaction to make the frequency integral in equation (10) well-behaved. This term is then added back into the self-energy:\cite{Shishkin2006}
\begin{equation}
\bar{W}_{\mathbf{q}}(\mathbf{G},\mathbf{G}^{\prime},\omega^{\prime})=W_{\mathbf{q}}(\mathbf{G},\mathbf{G}^{\prime},\omega^{\prime})-\nu^{\textrm{bare}}_{\mathbf{q}}(\mathbf{G},\mathbf{G}^{\prime})
\end{equation} 
\begin{equation}
\Sigma(\omega)_{n\mathbf{k},n\mathbf{k}}=\bar{\Sigma}(\omega)_{n\mathbf{k},n\mathbf{k}}+\langle\psi_{n\mathbf{k}}|\nu_{x}|\psi_{n\mathbf{k}}\rangle
\end{equation}

From equations (4) and (5) we see that the screened interaction $W(\mathbf{q})$ is bounded from above by the Hartree-Fock interaction, and thus in real systems ranges from small values, when the dielectric response $\epsilon(\mathbf{q})$ is large, up to this Hartree-Fock limit.

\subsection{Screening and Scattering}
Figure 1 lists some of the relevant Feynman diagrams for the processes under consideration in this study. Figure 1a illustrates a polarization bubble, which are summed over in the Random Phase Approximation (RPA) to generate the RPA self-energy (equations 7-10). It represents an electron in momentum state $\mathbf{k}$ emitting a photon of wavevector $\mathbf{q}$ at time \textit{t}, leaving it in momentum state $\mathbf{k-q}$. This photon is absorbed by state $\mathbf{p}$, promoting it to state $\mathbf{p+q}$, leaving a hole in state $\mathbf{p}$. At time \textit{t$^{\prime}$} state $\mathbf{p+q}$ emits a photon of wavevector $\mathbf{q}$, decaying back into state $\mathbf{p}$, which is absorbed by state $\mathbf{k-q}$ returning it to the original momentum state $\mathbf{k}$.

The RPA polarizability $\chi(\mathbf{q},\omega)$ gives the amplitude for bubbles of this type to occur (this is equation (7) without the overlap integrals):
\begin{equation}
\chi(\mathbf{q},\omega)=\sum\limits_{\mathbf{p}nn^{\prime}}\frac{f_{n^{\prime}\mathbf{p+q}}-f_{n\mathbf{p}}}{\omega-\epsilon_{n^{\prime}\mathbf{p+q}}+\epsilon_{n\mathbf{p}}\pm i\eta}
\end{equation} 

This will go ``on-shell" when $\omega-\epsilon_{n^{\prime}\mathbf{p+q}}+\epsilon_{n\mathbf{p}}\approx0$. For a metal with many states available near the Fermi level, $\epsilon_{n^{\prime}\mathbf{p+q}}-\epsilon_{n\mathbf{p}}\approx0$ for many transitions, and so $\omega$ is small. This corresponds to $t'-t\rightarrow\infty$, and therefore the polarization bubbles are long-lived, and the electrons are well screened. Taking this RPA interaction as a starting point (Assumption i), we can explore the effect of strong correlations by adding an interaction with another momentum state, $\mathbf{k^{\prime}}$.

If the transitions between the filled and the low-energy available states correspond to inter-site hopping (Assumption ii)), then the polarization bubble results in a double occupancy. If the Hubbard effective $U$ (equation (1)) is large with respect to the $t$ term, then the amplitude for another momentum state to interact with the bubble, and close it will be large. The first order approximation to this scattering process,\cite{Nozieres1962} $\Gamma(\mathbf{k},\mathbf{k^{\prime}})$, is illustrated in Figure 1b (exchange variants, in which each interaction $\nu(\mathbf{q})$ results in exchange are also possible but not pictured). The second interaction, $\nu^{\prime}(\mathbf{q})$, is given by:
\begin{equation}
\nu^{\prime}(\mathbf{q})=\frac{\int d^{3}\mathbf{r}\int d^{3}\mathbf{r^{\prime}}\phi^{*}_{\mathbf{p}}(\mathbf{r})\phi^{*}_{\mathbf{k^{\prime}+q}}(\mathbf{r^{\prime}})\phi_{\mathbf{p+q}}(\mathbf{r})\phi_{\mathbf{k^{\prime}}}(\mathbf{r^{\prime}})}{|\mathbf{r}-\mathbf{r^{\prime}}|}
\end{equation}

For interactions between momentum states constructed from $d$-orbitals, $|\mathbf{r}-\mathbf{r^{\prime}}|$ is small, and thus the interaction is strong. Therefore, this large amplitude will close the polarization bubble at some time \textit{t$^{\prime\prime}$} $<$ \textit{t$^{\prime}$}. The frequency at which the scattering vertex goes ``on-shell" shifts to higher $\omega$, which is illustrated in Figure 1c: the large amplitude for the interaction between $\mathbf{p+q}$ and $\mathbf{k^{\prime}}$ reduces the characteristic time the bubble stays open, shifting $t^{\prime\prime}$ closer to $t$.

This shift to higher frequency results in a stronger interaction between states $\mathbf{k}$ and $\mathbf{p}$ at longer times. Figure 1d illustrates this using a Hartree bubble (although an exchange interaction is also possible). Since the scattering process closes the bubble at time $t^{\prime\prime}$, a bare Coulomb interaction is now possible at time $t^{\prime\prime\prime}$, whereas without scattering the polarization bubble would still be open, and this interaction would be screened. Since the probability of the scattering vertex occurring is large due to the on-site overlap, diagrams such as Figure 1d may dominate those of Figure 1a, resulting in more interactions between the momentum states.

Computing the RPA dielectric function for CuO and M$_{2}$ VO$_{2}$ using a single $\mathbf{q}$ point, and comparing this to a calculation using the full $\mathbf{q}$ grid reveals that the dielectric response is dominated by the low $\mathbf{q}$ transitions (note that this does \textit{not} imply that the self-energy has a similar dependence), due to the photon propagator. The results are identical to the precision used in these calculations. Therefore, if the method used to generate the input wavefunctions is correct, but does not account for the on-site interaction (e.g. DFT), these low $\mathbf{q}$ transitions correspond to inter-site hopping, generating double occupancies. Since the dielectric response takes the form of a polarization bubble multiplied by a photon propagator, the creation of polarization bubbles which generate an energy penalty in the form of the Hubbard $U$ term will be found in the low $\mathbf{q}$ limit, and the scattering diagram which closes the bubbles will be the low $\mathbf{q}$ limit of Figure 1d, which is pictured in Figure 1e. From momentum conservation, if the photon propagator gives a $\sim 1/q^{2}$ dependency of the dielectric response, then the scattering vertex, $\Gamma(\mathbf{k},\mathbf{k^{\prime}})$, which ``corrects" the RPA screening, Figure 1b, must have the same $\mathbf{q}$ dependence at first order as the dielectric response. We make the ansatz that in the absence of significant Fermi surface nesting, the scattering which modifies the RPA and leads to Hubbard physics manifests in the low $\mathbf{q}$ limit. In addition, this suggests that in the limit of $\mathbf{q}\rightarrow 0$ the momentum states entering the scattering vertex are unchanged, i.e. forward scattering.

From the preceding argument, the consequences of on-site interactions from Mott systems will manifest as a change in the frequency dependence of the RPA screening. As mentioned in section I, the most common method of correcting this discrepancy is use the DFT+U approach, or a hybrid functional such as HSE06 which mixes in exact exchange to calculate the input wavefunctions. These shift the conduction band eigenvalues to higher energy, thus increasing the frequency of the polarization response. That is, these modify $\epsilon_{n^{\prime}\mathbf{k+q}}$, such that $\epsilon_{n^{\prime}\mathbf{p+q}}-\epsilon_{n\mathbf{p}} > 0$ (equation 13), increasing the frequency at which the process goes on-shell, and un-screening the low frequency interactions. However, there is another possibility. Rather than adjusting the eigenvalues, the frequency at which a bubble goes on-shell can be adjusted.  

Strong correlations in which electron localization is observed experimentally suggest that there is a considerable shift of the frequency dependence of the bubbles. Thus if the Hubbard $U$ term is large, corresponding to large amplitudes for the interaction at $t^{\prime\prime}$, the characteristic time of the polarization bubble will approach zero. This means that the Hartree-Fock interaction is effectively unscreened. A schematic of this is presented in Figure 1f, where the instantaneous closing of the polarization bubble is effectively a Hartree interaction (again, exchange is also a possibility), followed by further unscreened interactions. Thus the scattering vertex, (Figure 1b) which is a function of two single-particle Green functions, can be represented by one Green function. This reduces the computational load to that of a standard GW calculation. 

Therefore, if on-site interactions are generated by the low $\mathbf{q}$ limit of $\epsilon(\mathbf{q,\omega})$, then replacing the RPA response with the bare Hartree-Fock interaction for on-site interactions will simulate strong correlations. How this is achieved is detailed below.

\subsection{On-site Interactions in the Projector Augmented Wave Method}
In the PAW method\cite{Blochl1994b} utilized in the Vienna Ab Initio Simulation Package (VASP) code,\cite{Kresse1999} the all electron wavefunctions $|\psi_{n\mathbf{k}}\rangle$ are expanded as per:\cite{Gajdos2006}
\begin{equation}
|\psi_{n\mathbf{k}}\rangle =|\widetilde{\psi}_{n\mathbf{k}}\rangle+\sum\limits_{i}^{}(|\phi_{i}\rangle-|\widetilde{\phi}_{i}\rangle)\langle\widetilde{p}_{i}|\widetilde{\psi}_{n\mathbf{}k}\rangle
\end{equation}
The $|\widetilde{\psi}_{n\mathbf{k}}\rangle$ are the pseudowavefunctions, related to the cell periodic part of the wavefunctions through:
\begin{equation}
|\widetilde{\psi}_{n\mathbf{k}}\rangle =e^{i\mathbf{kr}}|\widetilde{u}_{n\mathbf{k}}\rangle
\end{equation}
and the $\widetilde{u}_{n\mathbf{k}}$ are expanded in plane waves. The partial waves $|\phi_{i}\rangle$, are solutions of the radial Schroedinger equation for a reference atom, while the pseudo-partial waves ($|\widetilde{\phi}_{i}\rangle$) are equivalent to the $|\phi_{i}\rangle$ outside a core radius $r_{c}$, and the projector functions are dual to the partial waves, $\langle\widetilde{p}_{i}|\phi_{j}\rangle=\delta_{ij}$. The subscript \textit{i} thus denotes the atomic position $\mathbf{R}_{i}$, angular momentum \textit{l$_{i}$} and \textit{m$_{i}$}, and the band index. Equations (7) and (10) require matrix elements of the form:
\begin{equation}
\langle\psi_{n^{'}\mathbf{k+q}}|e^{-i\mathbf{qr}}|\psi_{n\mathbf{k}}\rangle
\end{equation}
to be calculated. Following Gajdos \textit{et al.}\cite{Gajdos2006} this can be written:
\begin{equation}
\int B_{n^{\prime}\mathbf{k+q},n\mathbf{k}}(\mathbf{r})d^{3}\mathbf{r}
\end{equation}
where:
\begin{equation}
B_{n^{\prime}\mathbf{k+q},n\mathbf{k}}(\mathbf{r})\equiv e^{i\mathbf{qr}}\psi^{*}_{n^{\prime}\mathbf{k+q}}(\mathbf{r})\psi_{n\mathbf{k}}(\mathbf{r})
\end{equation}
Inserting the PAW expansion (equation (15)) this becomes:\cite{Gajdos2006}
\begin{multline}
B_{n^{\prime}\mathbf{k+q},n\mathbf{k}}(\mathbf{r})=\widetilde{u}^{*}_{n^{\prime}\mathbf{k}+{\mathbf{q}}}(\mathbf{r})\widetilde{u}_{n\mathbf{k}}(\mathbf{r})+\\\sum_{ij}^{}\langle \widetilde{u}_{n^{\prime}\mathbf{k}+{\mathbf{q}}}|\widetilde{p}_{i\mathbf{k+q}}\rangle\langle \widetilde{u}_{n\mathbf{k}+{\mathbf{q}}}|\widetilde{p}_{i\mathbf{k+q}}\rangle\\\times e^{i\mathbf{q}(\mathbf{r-R_{i}})}[\phi_{i}(\mathbf
r)\phi_{j}(\mathbf{r})-\widetilde{\phi}_{i}(\mathbf{r})\widetilde{\phi}_{j}(\mathbf{r})]
\end{multline}
In the low $\mathbf{q}$ limit the exponential is expanded as per:\cite{Gajdos2006}
\begin{equation}
e^{i\mathbf{q}(\mathbf{r-R_{i}})}=1+i\mathbf{q}(\mathbf{r-R_{i}})+o(\mathbf{q}^{2})
\end{equation}
The cell-periodic parts of the wavefunctions in $\mathbf{k}$-space are also expanded to first order around the valence wavefunctions:
\begin{equation}
\widetilde{u}_{n\mathbf{k}+{\mathbf{q}}}=\widetilde{u}_{n\mathbf{k}}+\mathbf{q}\nabla_{\mathbf{k}}\widetilde{u}_{n\mathbf{k}}+o(\mathbf{q}^{2})
\end{equation}
The low $\mathbf{q}$ limit of the exchange charge density then becomes:\cite{Gajdos2006}
\begin{equation}
\lim\limits_{\mathbf{q}\rightarrow 0}\langle\psi_{n^{'}\mathbf{k+q}}|e^{-i\mathbf{qr}}|\psi_{n\mathbf{k}}\rangle=|\mathbf{q}|\langle\hat{\mathbf{q}}\beta_{n'\mathbf{k}}|\widetilde{u}_{n\mathbf{k}}\rangle
\end{equation}
where $\hat{\mathbf{q}}=\mathbf{q}/|\mathbf{q}|$ and,\cite{Gajdos2006}
\begin{multline}
|\beta_{n\mathbf{k}}\rangle=(1+\sum\limits_{i}^{}|\widetilde{p}_{i\mathbf{k}}\rangle Q_{ij}\langle\widetilde{p}_{j\mathbf{k}}|)|\nabla_{\mathbf{k}}\widetilde{u}_{n\mathbf{k}}\rangle+\\i(\sum\limits_{ij}^{}|\widetilde{p}_{i\mathbf{k}}\rangle Q_{ij}\langle\widetilde{p}_{j\mathbf{k}}|(\mathbf{r}-\mathbf{R}_{i})|\widetilde{u}_{n\mathbf{k}}\rangle)-i(\sum\limits_{ij}^{}|\widetilde{p}_{i\mathbf{k}}\rangle \vec{\tau}_{ij}\langle\widetilde{p}_{j\mathbf{k}}|\widetilde{u}_{n\mathbf{k}}\rangle)
\end{multline}
with
\begin{equation}
Q_{ij}=\int_{\Omega(PAW)}^{}[\phi_{i}(\mathbf{r})\phi_{j}(\mathbf{r})-\widetilde{\phi}_{i}(\mathbf{r})\widetilde{\phi}_{j}(\mathbf{r})]d^{3}\mathbf{r}
\end{equation}
\\
\begin{equation}
\vec{\tau}_{ij}=\int_{\Omega(PAW)}^{}(\mathbf{r}-\mathbf{R}_{i})[\phi_{i}(\mathbf{r})\phi_{j}(\mathbf{r})-\widetilde{\phi}_{i}(\mathbf{r})\widetilde{\phi}_{j}(\mathbf{r})]d^{3}\mathbf{r}
\end{equation}

By setting $|\nabla_{\mathbf{k}}\widetilde{u}_{n\mathbf{k}}\rangle = 0$ in  equation (23), the planewave terms are eliminated, leaving only the augmentation sphere contributions. All of the cell periodic functions occur in overlap integrals with the projector functions, which are just the expansion coefficients for the all electron wavefunction (equation 15). The $\langle \widetilde{u}_{n^{\prime}\mathbf{k}+{\mathbf{q}}}|\widetilde{u}_{n\mathbf{k}}\rangle$ drop out due to orthogonality, and the only other cell-periodic terms are of the form $\langle \nabla_{\mathbf{k}} \widetilde{u}_{n^{\prime}\mathbf{k+q}}|...|\widetilde{u}_{n\mathbf{k}}\rangle$, which are set to zero. To see this more clearly, we can write the charge density at a point $\mathbf{r}$ the  for two orbitals $a$ and $b$ as\cite{Blochl1994b}
\begin{equation}
\psi^{*}_{a}(\mathbf{r})\psi_{b}(\mathbf{r})=n_{ab}(\mathbf{r})=\widetilde{n}_{ab}(\mathbf{r})-\widetilde{n}^{1}_{ab}(\mathbf{r})+n^{1}_{ab}(\mathbf{r})
\end{equation}
where $\widetilde{n}_{ab}(\mathbf{r})$ comes from the planewave expansion in equation (16), and the other two terms with the superscript 1, $\widetilde{n}^{1}_{ab}(\mathbf{r})$ and $n^{1}_{ab}(\mathbf{r})$ are \textit{one-center} terms coming from the $|\widetilde{\phi}_{i}\rangle$ and $|\phi_{i}\rangle$ respectively. They are only evaluated on the radial PAW grid,\cite{Shishkin2006} and thus only overlap when they correspond to the same atomic site, $\mathbf{R}_{i}=\mathbf{R}_{j}$. Setting $|\nabla_{\mathbf{k}}\widetilde{u}_{n\mathbf{k}}\rangle = 0$ in equation (23) is equivalent to eliminating the first term on the right-hand side of equation (27), leaving only the one-center terms. This reduction means that the interaction is non-zero only between the atomic-like wavefunctions inside the augmentation spheres on the same site, \textit{i.e.} the interaction is \textit{on-site}.

In the implementation of the GW approximation used in this study,\cite{Shishkin2006} the dielectric matrix of equation (8) is set equal to the unit diagonal whenever terms inside atomic spheres are evaluated, resulting in a bare Hartree-Fock interaction. Thus, by eliminating the plane wave terms in the low $\mathbf{q}$ limit, a very strong (Hartree-Fock) on-site only interaction replaces the screened interaction in the self-energy. Unscreening the interaction in this manner will therefore simulate the behavior of the scattering vertex of Figure 1b, as the polarization bubbles are largely, although not completely, eliminated as is explained below.

From equations (18) and (19) it is also clear that the magnitudes of the bare Coulomb terms are controlled in part by the derivatives of the cell periodic parts of the wavefunctions, arising from the first order expansion (17). These derivatives are evaluated in gapless systems using second order perturbation theory, which creates issues for self-consistency, as once the first round of quasiparticle shifts are evaluated the DFT Hamiltonian is no longer valid. Such gapless systems also include Mott insulators if DFT is used to generate the input wavefunctions. It is obvious however, that setting these terms to zero, $|\nabla_{\mathbf{k}}\widetilde{u}_{n\mathbf{k}}\rangle = 0$, will both remove this obstacle to self-consistency, and reduce the magnitudes of the overlap integrals which weight the interaction $\langle\psi_{n\mathbf{k}}|\nu_{x}|\psi_{n\mathbf{k}}\rangle$. Therefore, the removal of the $|\nabla_{\mathbf{k}}\widetilde{u}_{n\mathbf{k}}\rangle$ terms from the calculation has two effects. They both completely unscreen the low $\mathbf{q}$ interactions, by setting the dielectric matrix to the unit diagonal, and also reduce the magnitudes of the bare Coulomb terms added into the self-energy. This effectively replaces the screened interaction at low $\mathbf{q}$ with a stronger interaction, however in which small amount of screening is still present. Note that when the self-energy is evaluated as per equation (10), the full PAW wavefunctions are used, which are orthonormal.

The limitation of this approach to long range interactions in this work renders the method applicable only to low symmetry Mott systems which do not exhibit significant Fermi surface nesting at finite wavevector. If nesting is present, then the polarizability $\chi(\mathbf{q})$ will exhibit a peak at the nesting wavevector, corresponding to a significant amplitude for particle-hole formation. Since our method does not penalize pair bubbles at finite $\mathbf{q}$, the interactions will be underestimated by the RPA correlations, as usual. However, despite this, the results for low symmetry systems are in general quite illuminating.

\subsection{Computational Details}
The computational scheme used for CuO was as follows. The experimentally determined structural parameters\cite{Wyckhoff1963} were used as input to density functional theory\cite{Kohn1965,Kresse1996} calculations on $8\times 8\times 6$ Monkhorst-Pack k-space grids, using the Generalized Gradient Approximation approach to exchange and correlation of Perdew \textit{et al.} (PBE),\cite{Perdew1996} and the Brillouin zone integration approach of Bloechl \textit{et al.}\cite{Bloechl1994} No initial spin ordering was assumed in all calculations. GW calculations were performed using the implementation of Shishkin and Kresse\cite{Shishkin2006,Shishkin2007} in the Vienna Ab Initio Simulation Package (VASP)\cite{Kresse1996} in either fully frequency dependent, or static ($\omega$ = 0) modes using 256 bands. The frequency dependent calculations were performed as one-shot $G_{0}W_{0}$ calculations, while the Partially-Screened GW calculations utilized three self-consistency iterations ($G_{3}W_{3}$). An energy cutoff of 200 eV was used for all GW calculations.

The M$_{1}$ and M$_{2}$ structures used were those of Andersson\cite{Andersson1954} and Marezio \textit{et al.}\cite{Marezio1971} respectively. The M$_{1}$ structure was first relaxed to a ground state using GGA DFT (PBE).\cite{Perdew1996} The M$_{2}$ structure was not relaxed, as DFT underestimates the correlation energy,\cite{Eyert2002} and thus reaches an incorrect ground state. $6\times 6\times 6$ and $4\times 6\times 6$ Monkhorst-Pack k-space grids were used for the M$_{1}$ and M$_{2}$ structures respectively and the GW calculations were performed again with VASP using 256 bands, after first calculating input wavefunctions using DFT with PBE GGA functionals. The static PS-GW calculations of the M$_{2}$ structure utilized four self-consistency steps ($G_{4}W_{4}$) and 256 bands.

Convergence tests for both CuO and M$_{2}$ VO$_{2}$ are presented in the Appendix. For both systems convergence was achieved in a relatively small number of self-consistency steps; 3 for CuO and 4 for M$_{2}$ VO$_{2}$.

\begin{figure*}[th!]
  \includegraphics[width=1.5\columnwidth]{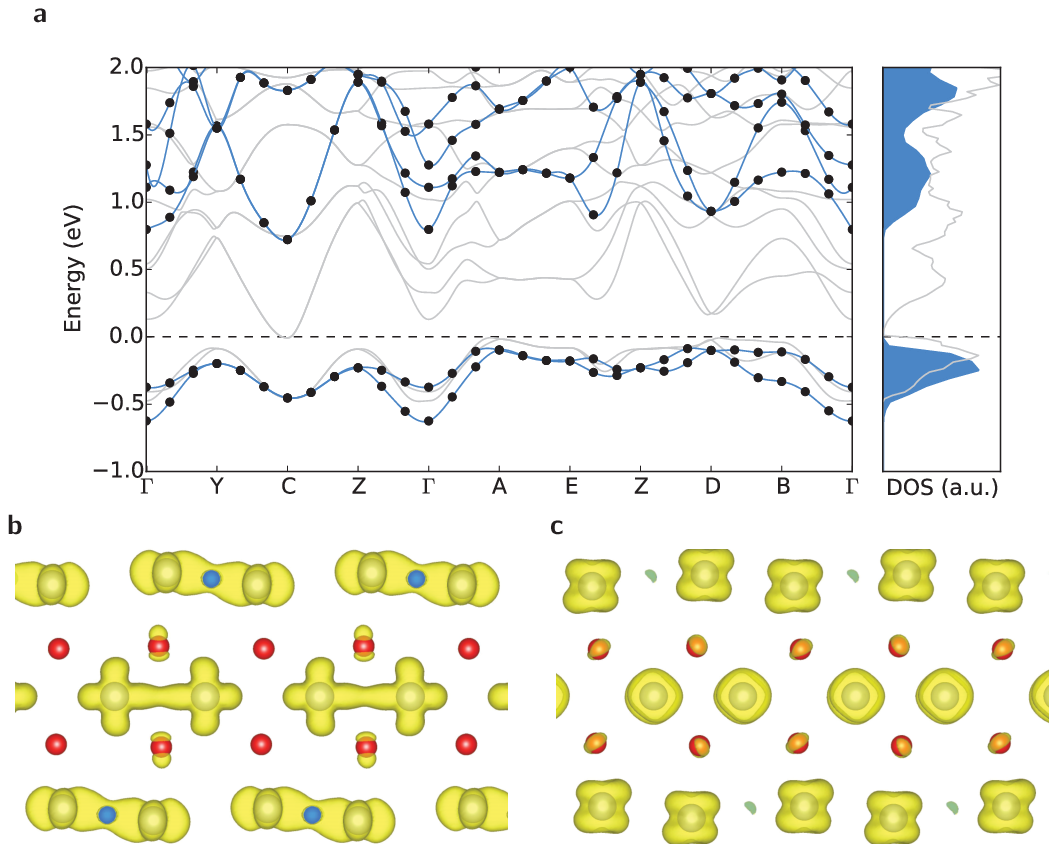}
  \caption{a) Left panel: M$_{1}$ VO$_{2}$ band structures calculated using DFT (gray lines) and unmodified frequency-dependent $G_{0}W_{0}$ (black filled circles fitted with blue splines), right panel: corresponding DFT (gray line) and $G_{0}W_{0}$ (blue filled curve) densities of states, b) charge density isosurface of the valence band of M$_{1}$ VO$_{2}$, and c) charge density isosurface of the conduction band of M$_{1}$ VO$_{2}$. Both isosurfaces perspectives correspond to the (0$\bar{1}$1) plane.} 
\end{figure*}

\section{Results and discussion}
\subsection{Application to CuO}
Figure 2 details the application of this technique to the Mott insulator CuO.\cite{Takahashi1997} Figure 2a illustrates the density of states (DOS) near the Fermi level of a Hartree-Fock calculation, and as is commonly observed, while the Hartree-Fock approach does open a gap, it is overestimated, and predicts the ground state to be Ferromagnetic, again contradicting experimental data.  Figure 2b presents a standard (unmodified) spin-resolved $G_{0}W_{0}$ calculation of the states at the Fermi level. As is expected from the independent particle-RPA approach, the non-interacting Green functions and neglect of scattering vertices over-screens the Hartree-Fock interaction, resulting in metallic behavior. In fact, very little difference is exhibited between the $G_{0}W_{0}$ and DFT calculations (a comparison of the DFT and standard $G_{0}W_{0}$ band structures is presented in the Supporting Information). PS-$G_{3}W_{3}$ calculations however reveal a different story (Figure 2c). Clear splitting of the states at the Fermi level into two characteristic peaks, the upper and lower Hubbard bands (UHB and LHB labels on Figure 2c) is exhibited (although the use of the static limit does broaden the lower lying oxygen bands, see Supporting Information), and an excitation gap of approximately 1.1 eV has opened. 

\begin{figure*}[th!]
  \includegraphics[width=1.8\columnwidth]{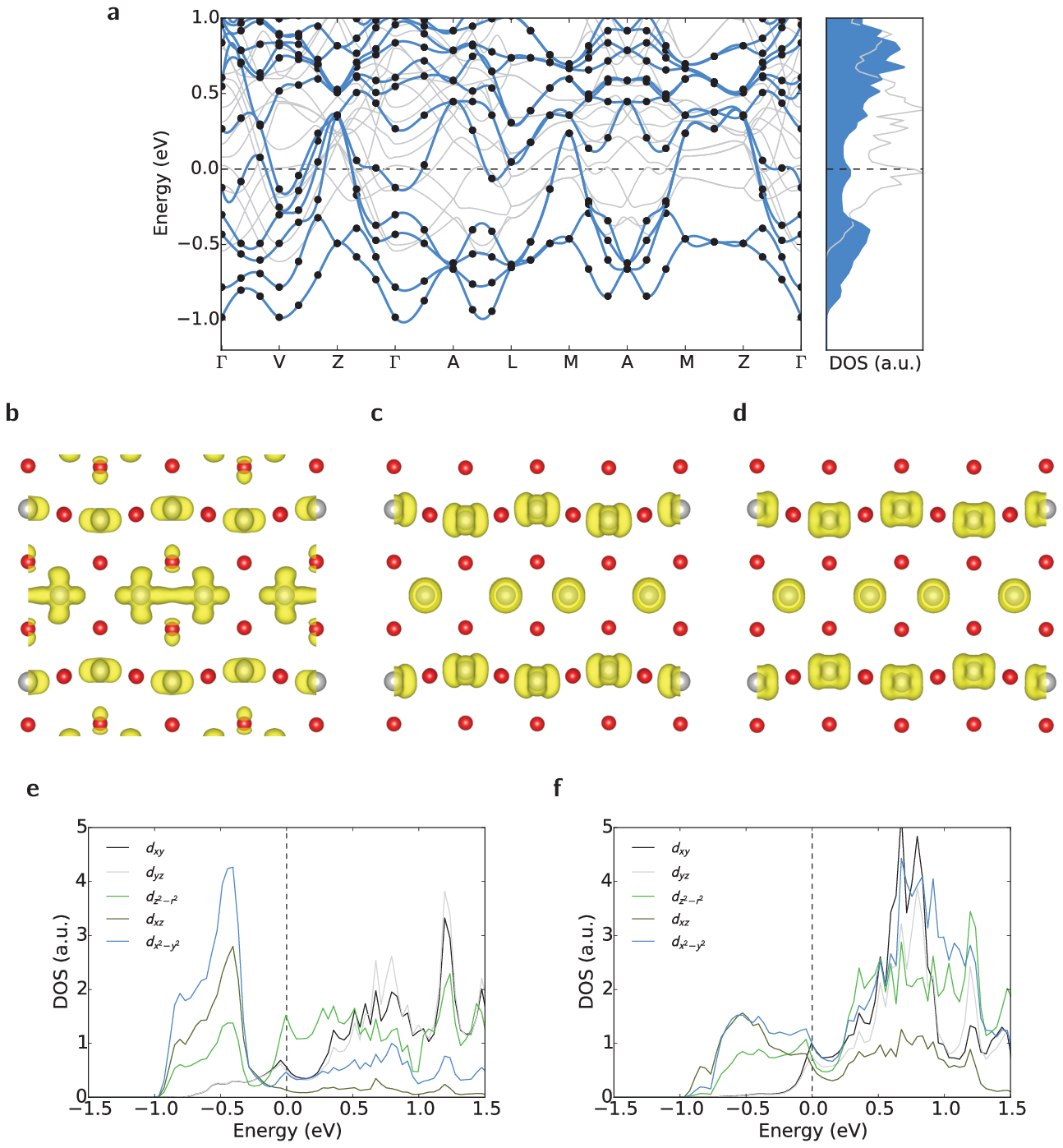}
  \caption{a) Left Panel: M$_{2}$ VO$_{2}$ band structures calculated using DFT (gray lines) and unmodified frequency-dependent $G_{0}W_{0}$ (black filled circles, fitted with blue splines), right panel: corresponding DFT (gray line) and $G_{0}W_{0}$ (blue filled curve) densities of states,  b) charge density isosurface of the valence band of M$_{2}$ VO$_{2}$, c) charge density isosurface of the quasiparticle peak of M$_{2}$ VO$_{2}$, d) charge density isosurface of the conduction band of M$_{2}$ VO$_{2}$, e) projected densities of states of the Peierls chain vanadium atoms of M$_{2}$ VO$_{2}$, f) projected densities of states of the antiferroelectric chain vanadium atoms of M$_{2}$ VO$_{2}$.}
\end{figure*}

The magnitude of this gap compares relatively favorably with the experimentally determined values of 1.4-1.7 eV, \cite{Heinemann2013} and the overall shape of the DOS is in reasonable agreement with R{\"o}dl \textit{et al.}\cite{Claudia2015} apart from the aforementioned broadening. However there is another experimental signature of Mott systems which must be simulated in order for the technique to be considered an accurate reproduction. When doped with electrons or holes, Mott systems exhibit a “giant transfer of spectral weight” which clearly illustrates the failure of band theory for these systems.\cite{Eskes1991} In a system well-described by band theory, if an electron is doped into the conduction band, the Fermi level shifts up, and the conduction band intersects the Fermi level with minimal change in dispersion. In a Mott system, adding a small number of carriers is not expected to significantly affect the $t/U$ balance, and the system is still expected to be gapped. Therefore, any previously empty state  in the upper Hubbard band which is filled upon doping must then cross the gap to sit in the lower Hubbard band, which significantly changes band dispersion. This effect was clearly observed in recent photoemission experiments on TiOCl.\cite{Sing2011} Figure 2d presents static PS-$G_{3}W_{3}$ calculations of CuO doped with one electron. When compared to Figure 2c, it is clear that spectral weight has shifted from the upper Hubbard band to sit at the leading edge of the lower Hubbard band. Figures 2e-f illustrate how this occurs. Figure 2e is a comparison of the DFT and static PS-$G_{3}W_{3}$ band structure calculations. We use DFT for this comparison rather than standard GW data as the standard GW data is virtually identical to the DFT data (see Supporting Information), and the lower computational cost of DFT allows us to use much higher $\mathbf{k}$-space resolution. Figure 2f presents the same data for the electron doped structure. The calculations of the undoped structure reveal that the effect of unscreening the low $\mathbf{q}$ interactions is to split the spectral weight at the Fermi level. As noted in numerous DMFT calculations, band crossings at the Fermi level still exist. \cite{Aichhorn2009,Aichhorn2010,Ferber2012,Werner2012} Electron doping (Figure 2f) fills the lowest lying states in the upper Hubbard band, which Figure 2e indicates are around the $\Gamma$ point, dropping them onto the leading edge of the lower Hubbard band, at approximately -1 eV.

\begin{figure*}[th!]
  \includegraphics[width=1.5\columnwidth]{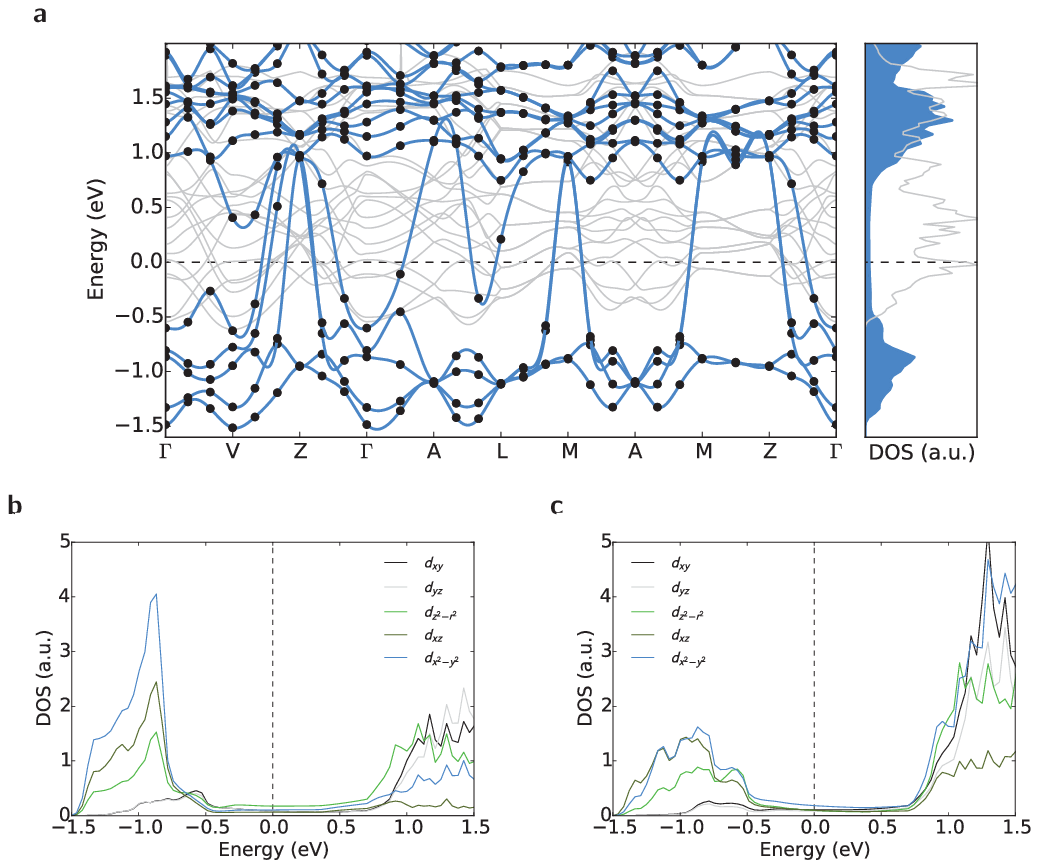}
  \caption{a) Left Panel: M$_{2}$ VO$_{2}$ band structures calculated using DFT (gray lines) and static PS-$G_{4}W_{4}$ (black circles, fitted with blue splines), right panel: corresponding DFT (gray line) and static PS-$G_{4}W_{4}$ (blue filled curve) densities of states, b) static PS-$G_{4}W_{4}$ projected densities of states of the Peierls chain vanadium atoms of M$_{2}$ VO$_{2}$ and c) static PS-$G_{4}W_{4}$ projected densities of states of the antiferroelectric chain vanadium atoms of M$_{2}$ VO$_{2}$.}
\end{figure*}

\subsection{Application to Vanadium Dioxide}
Turning now to the technologically significant problem of the natures of the insulating phases of vanadium dioxide \cite{Nakano2012,Liu2012,Park2013,Budai2014}, Figure 3 presents $G_{0}W_{0}$ calculations on the M$_{1}$ form. This structure undergoes an insulator-metal transition at $\sim$ 340 K as it spontaneously changes from the monoclinic P2$_{1}$/c structure to the tetragonal P4$_{2}$/mnm form. \cite{Goodenough1971} Figure 3a presents a comparison of the DFT and $G_{0}W_{0}$ bands, and the respective densities of states and the data clearly illustrates that the $G_{0}W_{0}$ calculations result in splitting of the bands with respect to the DFT calculation, with the empty conduction band simply shifting upwards, with minimal change in dispersion. The gap magnitude of $\sim$ 0.7 eV is in excellent agreement with the experimental value, which is also $\sim$ 0.70 eV\cite{{Shin1990}} and the scCOHSEX-G$_{0}$W$_{0}$ calculations of Gatti \textit{et al.}\cite{Gatti2007}

Figures 3b-c present charge density isosurfaces of the valence (Fig. 3a) and conduction (Fig. 3c) bands in the ($\bar{1}$,$1$,$0$) plane (the ``conduction band" is the first peak above the Fermi level in the Density of States).  As expected from the well-known Peierls distortion of the M$_{1}$ structure, the pairing of the vanadium nuclei results in bonding density between the nuclei, while the conduction band consists of the corresponding anti-bonding states, thus confirming that the gap in M$_{1}$ VO2 opens via bonding/antibonding splitting. The magnitude is significantly underestimated by DFT however. Figures 3b-c indicate that Peierls pairing produces an increase in the inter-vanadium local potential, which stabilizes bonding wavefunctions with respect to conduction states, shifting the conduction band eigenvalues to higher energy.

Figure 4 illustrates that this is not the case for the M$_{2}$ form. M$_{2}$ vanadium dioxide also undergoes an insulator-metal transition, although at slightly higher temperature (353 K \cite{Booth2009}), coincident with a structural transition from monoclinic C2/m to the same tetragonal P4$_{2}$/mnm structure as the M$_{1}$ form.\cite{Pouget1974} However, the monoclinic form differs significantly in structure. In the M$_{1}$ form all of the vanadium atoms form Peierls paired chains running down the monoclinic a-axis, which experience a slight antiferroelectric twist that has components in both the b- and c-axes. The M$_{2}$ form however, has two distinct chain structures (a comparison of this with the M$_{1}$ structure is presented in the Supporting Information). One half of the vanadium atoms form a Peierls paired chain, however this chain is not antiferroelectrically distorted, but rather the vanadium atoms are collinear. The remaining vanadium atoms form an antiferroelectrically distorted chain, however one in which the inter-vanadium spacing is uniform (\textit{i.e.} no Peierls pairing). This structure has been regarded as a Mott insulator since the 1970s, due to experiments by Pouget and co-workers\cite{Pouget1974} who used $^{51}$V NMR Knight shifts to resolve the two vanadium environments in Cr-doped VO$_{2}$. Figure 4a presents a comparison of a standard $G_{0}W_{0}$ calculation of the M$_{2}$ structure using a grid of 30 frequency points with DFT data. While some splitting of the DOS at the Fermi level is evident in comparison to DFT, a peak is still observed at $E_{F}$. The $G_{0}W_{0}$ band structure confirms that the while there is splitting of the bands, states are evident at the Fermi level in the $Z-\Gamma-A$ directions and at $L$.  This splitting is suggestive of the lower Hubbard band-quasiparticle peak-upper Hubbard band splitting\cite{Kotliar2006} observed in DMFT studies of correlated metals such as paramagnetic V$_{2}$O$_{3}$,\cite{Mo2003} however closer inspection reveals a more practical way to regard these features. 

Given that the structure contains a Peierls chain, it is expected that there will be some bonding-antibonding splitting observed, as per the M$_{1}$ structure. However, as this chain does not undergo antiferroelectric distortion, the vanadium and oxygen orbitals along the $z$-axis of the chain are not Peierls paired. Therefore a non-bonding, metallic band is expected to exist at $E_{F}$. Figures 4b-d illustrate this with charge density isosurfaces of the lower valence band (Figure 4b), the “quasiparticle” band (Figure 4c), and the conduction band (Figure 4d). Clearly, the valence band contains all of the bonding density, while the quasiparticle and conduction bands are non-bonding/antibonding. Projecting the density of states onto atomic-like orbitals on the Peierls chain vanadium atoms (Figure 4e) illustrates that the quasiparticle peak is indeed mostly non-bonding $3d_{z^{2}-r^{2}}$ states. The antiferromagnetic (AF) chain (Figure 4f) in contrast exhibits more mixed character at $E_{F}$. Thus, Figure 4 indicates that in contrast to the M$_{1}$ structure, standard $G_{0}W_{0}$ calculations predict that the M$_{2}$ structure is metallic due to the reduced bonding/antibonding splitting brought about by the change in structure. This is at odds with the experimentally determined insulating behavior.

Static PS-$G_{4}W_{4}$ calculations however suggest far more localised behaviour. The density of states at $E_{F}$ (Figure 5a) is considerably reduced, with the spectral weight splitting in a fashion similar to the CuO data of Fig 2c, with some states moving down to form a broad lower Hubbard band with the Peierls bonding states, while some move upwards into the antibonding band, creating an upper Hubbard band, separated from the lower Hubbard band by a gap of $\sim$ 1 eV. The band structure (Figure 5a) reveals considerable depletion of the states at the Fermi level, with the valence states shifting downwards as the gap opens, and the oxygen $p$-states at $\sim$ -3 to -4 eV shift concurrently. The projected DOSs for the Peierls chain (Figure 5b) illustrates that the non-bonding $3d_{z^{2}-r^{2}}$ states observed in the $G_{0}W_{0}$ calculation have disappeared from the gap, as have the mixed $d$-states of the AF chain (Figure 5d). It is evident that increasing the effect of electron correlations splits the metallic non-bonding states into Hubbard bands, as expected of a Mott insulator.

\section{Conclusion}
In summary, accounting for scattering due to on-site repulsion in the low $\mathbf{q}$ limit of the polarizability in GW calculations of Mott insulators correctly reproduces some of the signatures of Mott physics. Specifically, for low-symmetry structures in which significant Fermi surface nesting is not expected, the approach generates the splitting of the states at the Fermi level into the upper and lower Hubbard bands, and the giant transfer of spectral weight with electron doping. Upon application of this extension to the M$_{2}$ form of VO$_{2}$ it is evident that the M$_{1}$ and M$_{2}$ forms are fundamentally different types of insulator. The M$_{1}$ form opens a gap from bonding anti-bonding splitting; the bonding valence band wavefunctions are stabilized with respect to the valence band states. This reduces polarizability and unscreens the conduction band, giving rise to a correlated band insulator.\cite{Belozerov2012} This type of insulator is well described by conventional $G_{0}W_{0}$ calculations. The M$_{2}$ structure however opens a gap due to strong local \textbf{k}-space correlations, it is a Mott insulator.\cite{Mott1974} 

The approach described however rests upon the assumption that after construction of the input wavefunctions and eigenvalues by DFT, the low \textbf{q} dielectric response contains all of the polarizations that create double occupancies, and nothing else. While mathematically there is certainly a low \textbf{q} \textit{dependence} due to the photon propagator, this is not a rigorous enough, or general enough approach to be widely applicable. However, the computation of the response function is far less taxing than the self-energy, and the possibility exists to implement a more rigorous approach to the determination of transitions which generate double occupancies. If such an approach could be found, modified self-energy calculations of the kind presented here could provide significant insight into the nature of Mott insulating materials from an \textit{ab initio} perspective, and thus  significantly facilitate materials design.

\section{Appendix}
Figures 6a-c present convergence tests using the total Density of States for CuO spin-resolved (Figure 6a) and spin-independent (Figure 6b) calculations, and spin-independent M$_{2}$ VO$_{2}$ calculations (Figure 6c). In all cases convergence for static calculations (\textit{i.e.} those used in this study) occurred quickly. For CuO, three iterations was sufficient, while for M$_{2}$ VO$_{2}$ convergence was reached within four iterations.
\begin{figure}
  \includegraphics[width=0.8\columnwidth]{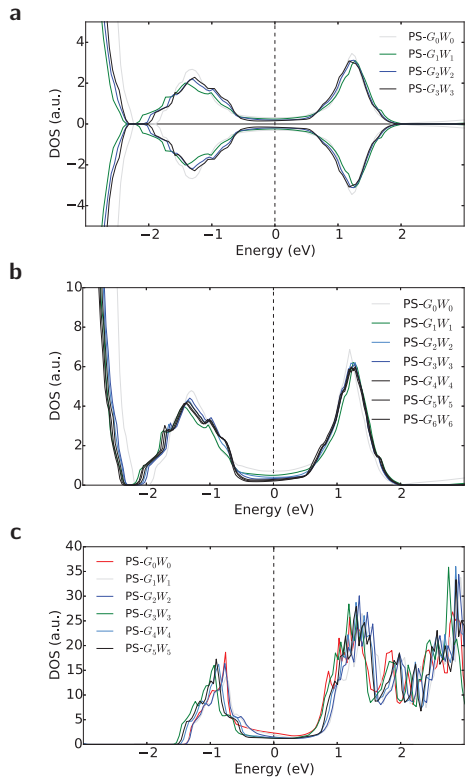}
  \caption{a) Effect of self-consistency iterations on the spin-resolved PS-GW data for CuO, b) effect of self-consistency on the spin-independent and PS-GW data for CuO (the lower computational requirements allow more iterations to be used), the data for 4, 5 and 6 iterations are all plotted in black as they practically overwrite each other, and c) effect of self-consistency iterations for the M$_{2}$ VO$_{2}$ structure.}
\end{figure}

\section{Acknowledgements}
This work was supported by computational resources provided by the Australian Government through the National Computational Infrastructure under the National Computational Merit Allocation Scheme. DWD acknowledges the support of the ARC Centre of Excellence for Nanoscale BioPhotonics (CE140100003).
\section{Correspondence}
Correspondence and requests for materials should be addressed to J.M.B., email: jamie.booth@rmit.edu.au

\section{Supplementary Information}

Illustrations of the structures of CuO, M$_{1}$ and M$_{2}$ vanadium dioxide are presented in the supporting information. Also included is an exploration of the effects of frequency dependence on the $d$- and $p$-states below the Fermi level of CuO. A comparison of the standard G$_{0}$W$_{0}$ and DFT band structures of CuO, and the effect of changing the resolution of the \textbf{k}-point grid is also presented.

\bibliography{library}
\end{document}